\documentclass[12pt]{article}

\usepackage{epsfig,amsmath,amssymb,graphics,graphicx} 

\usepackage{amscd}
\usepackage{amsfonts}

\usepackage{amsmath,amsfonts,amsthm,amssymb,eucal}
\usepackage{breqn}

\newtheorem{Definition}{Definition}[section]
\newtheorem{Theorem}{Theorem}[section]
\newtheorem{Lemma}{Lemma}[section]

\begin{document}

\begin{center}

Computational and Applied Mathematics 40 (2021) 296

\vskip 7mm

{\bf \large From Fractional Differential Equations with Hilfer Derivatives}
\vskip 3mm
{\bf \large To Discrete Maps with Memory} \\

\vskip 7mm
{\bf \large V.E. Tarasov}$^{1,2}$ \\
\vskip 3mm

${}^1$ {\it 
Lomonosov Moscow State University, Moscow 119991, Russia} \\
${}^2$ {\it Moscow Aviation Institute (National Research University), Moscow 125993, Russia } \\

\begin{abstract}
In this article, we proposed new discrete maps with memory (DMM). 
These maps are derived from fractional differential equations (FDE) with the Hilfer fractional derivatives of non-integer orders and periodic sequence of kicks. 
The suggested DMM are obtained from these equations without any approximation, and they are a discrete form of the exact solutions. 
DMM are proposed for arbitrary positive orders of 
equations with the Hilfer fractional derivatives. 
As an example, the suggested maps are described for the orders lying in the intervals (0,1) and (1,2). 
The maps, which are derived from the equations with the Hilfer operators, allow us to consider a whole range of the maps derived from the FDE with the Caputo and Riemann-Liouville fractional derivatives.
\end{abstract}

\end{center}

\noindent

\noindent


\section{Introduction}

Fractional differential equations (FDE) 
are a special type of integro-differential equations.
The integro-differential operators, which are used in this type of equations, form a calculus that is called fractional calculus (see \cite{FC1,FC4,Handbook2019-1,Handbook2019-2}).
FDE are used to describe processes and systems with 
non-locality in time and fading memory in 
physics \cite{Handbook2019-4,Handbook2019-5}, 
economics \cite{BOOK-DG-2021}, 
biology \cite{Ionescu}, and other sciences. 

In nonlinear dynamics and theory of deterministic chaos, discrete maps may be derived from differential equations of integer orders with periodic kicks (see Section 5 in \cite{SUZ},
Sections 5.2 and 5.3 in \cite[pp.60-68]{Zaslavsky2}, and Section 1.2 in \cite[pp.16-17]{Schuster}, Chapter 18 in \cite[pp.409-453]{Springer2010}). 
All these discrete maps are memoryless maps, since values of map variables at the next step are determined only by 
their values at the previous step, and it can be represented in the form $Z_{n+1}=f(Z_n)$. 
In discrete maps, the memory means that the present values of variables depend on all their past values, i.e. $Z_{n+1}=f(Z_n,Z_{n-1}, ...,Z_1)$. 

For the first time, discrete maps with memory (DMM) were obtained from fractional differential equations (FDE) with Riemann-Liouville and Caputo fractional operators in works 
\cite{Tarasov-Zaslavsky,Tarasov-Map1,Tarasov-Map2} 
(see also \cite[pp.409-453]{Springer2010} and 
\cite{Tarasov-Map3,TT-Logistic,Entropy2021,MMAS2021}).
In the proposed approach, the DMM are derived from exact solutions of FDE without any approximations
\cite[pp.409-453]{Springer2010}).
Then, this approach has been applied to describe properties of the DMM (for example, see \cite{Tarasov-Map3,TT-Logistic,Entropy2021,MMAS2021,TT-Edelman1,TT-Edelman2,Edelman3}, and
\cite{Edelman-Handbook-2,Edelman-Handbook-4,Edelman-2021}. 
New types of attractors and chaos were 
were discovered in computer simulations. 

Let note the importance of discrete map with nonlocality in time for economics \cite{TT-Logistic}, quantum physics \cite{Entropy2021} and population dynamics \cite{MMAS2021}.
The DDM can be obtained as exact solutions for systems that are described by the fractional integral equations \cite{Mathematics2021}. 

It should be emphasized that nonlocality in time in discrete maps can be considered not only for systems with memory, but also for systems with nonlocal scaling \cite{CNSNS2021,CSF2021}, and for general fractional dynamics with general form of nonlocality in time \cite{Mathematics2021-2} including distributed time delay.
For discrete maps with general form of nonlocality in time, the Edelman methods \cite{Edelman-2021} can be used to find new types of attractors and chaotic behavior. 


In the proposed paper, new DMM are derived from FDE with Hilfer fractional derivatives. 
The Hilfer fractional operators was suggested in 2000 \cite[p.~113]{Hilfer2000}). 
The fundamental theorems of fractional calculus for the Hilfer fractional derivatives are described in article \cite{HilferLuchkoTomovski2009}.
Then, these operators were generalized \cite{GGPT2014,SO2018}. 
The well-known Riemann-Liouville and Caputo fractional derivatives \cite{FC1,FC4} are special cases of the Hilfer fractional derivative (HFD), when the parameter $\nu \in [0,1]$ is equal to 0 and 1, respectively.

In this paper, we proposed nonlinear FDE with Hilfer fractional derivatives and kicks. 
Solutions of these FDE are derived for arbitrary orders of fractional derivatives.
Using these exact solutions, we derive DMM without using any approximations. 
As an example, the proposed universal maps with memory are described for the orders from (0,1) and (1,2).


\section{Equations with Hilfer FD and Kicks}

We begin by defining the function spaces, which are used to formulate the fundamental theorems of fractional calculus with the Hilfer fractional derivatives 
\cite{HilferLuchkoTomovski2009}. 

\begin{Definition}
Function $Z(t)$ belongs to the space $C_{\gamma}$, $\gamma \in \mathbb{R}$, if there exists a real number $p > \gamma$, such that $Z(t) = t^p \, Y(t)$, where $t>0$ and $Y(t) \in C[0,\infty)$.
\end{Definition}

\begin{Definition}
Function $Z(t) \in C_{-1}$ belongs to 
the space $\Omega^{q}_{-1}$, $q \ge 0$, 
if $(D^{\mu,\nu}_{H,0+}Z)(t) \in C_{-1}$
for all $0 \le \mu \le q$, $0 \le \nu \le 1$.
\end{Definition}

Note that $\Omega^{0}_{-1}=C_{-1}$.
The space $\Omega^{\mu}_{-1}$ contains in particular all functions $Z(t) =t^{\gamma} Y(t)$ with $\gamma \ge \mu$, where $Y(t)$ is an analytical function on $[0,\infty)$.

The Hilfer FD was suggested in \cite{Hilfer2000} in 2000 (see Definition 3.3 in book \cite[p.~113]{Hilfer2000}).

\begin{Definition}
For $Z(t) \in \Omega^{\mu}_{-1}$, $N-1<\mu\le N$, and $\nu \in [0,1]$, 
the Hilfer fractional derivative (HFD) of the order $\mu$ and type $\nu$ is defined by the equation 
\begin{equation} \label{Hilfer-Hilfer-1} 
\left(D^{\mu,\nu}_{H;0+} Z\right)(t) =
\left( I^{\nu (N-\mu)}_{RL;0+}\frac{d^N}{dt^N} 
I^{(1-\nu) (N-\mu)}_{RL;0+} Z \right)(t) ,
\end{equation} 
where 
\begin{equation} \label{EQ1-27} 
\left(I^{\mu}_{RL;0+} Z \right)(t) = \frac{1}{\Gamma (\mu)}\int^t_0 (t-\tau)^{\mu-1} Z (\tau) d\tau .
\end{equation} 
Here $\Gamma (\mu)$ is the gamma function, and $Z(t) \in L_1(0,b)$.
\end{Definition}

For operator \eqref{Hilfer-Hilfer-1}, the parameter $\nu \in [0,1]$ gives a possibility to consider a whole range of fractional derivatives between the the Riemann-Liouville fractional derivative $D^{\mu}_{RL;0+}$ and the Caputo fractional derivative $D^{\mu}_{C;0+}$ such that
\begin{equation} \label{Hilfer-Hilfer-1b} 
\left(D^{\mu,0}_{H;0+} Z\right)(t) = 
\left(D^{\mu}_{RL;0+} Z\right)(t) , \quad
\left(D^{\mu,1}_{H;0+} Z\right)(t) = 
\left(D^{\mu}_{C;0+} Z\right)(t) .
\end{equation} 

Let us consider the fractional differential equation
\begin{equation} \label{FDE-1}
\left(D^{\mu,\nu}_{H;0+} Z\right)(t) + \Phi[Z(t)] \sum^{\infty}_{k=1} \delta (T^{-1}t-k) = 0 ,
\end{equation}
in which perturbation is a periodic sequence of kicks, $T$ is a period, $\Phi [Z]$ is a real-valued function, 
$D^{\mu,\nu}_{H;0+}$ is the HFD of the order $\mu \in [N-1, N]$ and type $\nu \in [0,1]$.

The product of the generalized function $\delta (t/T-k)$, and the functions $\Phi [Z(t)]$ is defined if $\Phi [Z(t)]$ is continuous for $t=kT$. 
Therefore, we will use \cite{Edelman3} equation \eqref{FDE-1} in the form
\begin{equation} \label{FDE-1e}
\left(D^{\mu,\nu}_{H;0+} Z\right)(t) + \Phi[Z(t - \varepsilon)] \sum^{\infty}_{k=1} \delta (T^{-1}t-k) = 0 .
\end{equation}


To derive DMM from FDE \eqref{FDE-1e}, we will use the second fundamental theorem of fractional calculus in the following form.

\begin{Theorem} \label{FTFC-Hilfer-1}
Let $Z(t) \in \Omega^{\mu}_{-1}$, $N-1 \le \mu \le N$, $N \in \mathbb{N}$, $0 \le \nu \le 1$.
Then, the operators $I^{\mu}_{RL,0+}$ and $D^{\mu,\nu}_{H;0+}$ are connected 
\begin{equation} \label{FTFC-Hilfer-2} 
\left(I^{\mu}_{RL,0+} D^{\mu,\nu}_{H;0+} Z\right)(t) =
Z(t)- \sum^{N-1}_{m=0} 
\frac{p^{\mu,\nu}_m(0+)}{\Gamma(m+1- (N-\mu)(1-\nu))}
\, t^{m - (N-\mu)(1-\nu)} ,
\end{equation} 
where
\begin{equation} \label{p-k0-1}
p^{\mu,\nu}_m(0+) = \lim_{t \to 0+} \frac{d^m}{dt^m} 
\left( I^{(1-\nu) (N-\mu)}_{RL;0+} Z \right)(t) ,
\end{equation} 
where $0 \le \nu \le 1$, $t>0$, $m=0,...,N-1$. 
\end{Theorem}

\textbf{Proof.}
Theorem \ref{FTFC-Hilfer-1} was proved in paper as Theorem 2 in \cite[p.~304]{HilferLuchkoTomovski2009}. 

$\ \ \ \Box$ \\

Using Theorem \ref{FTFC-Hilfer-1}, we can prove the following statement that describes the exact solution of equation \eqref{FDE-1e}.

\begin{Theorem} \label{Hilfer-Th-Zt-1}
Let $Z(t) \in \Omega^{\mu}_{-1}$, $N-1 < \mu <N$, $\nu \in [0,1]$. 
The Cauchy problem for the equation
\begin{equation} \label{FDE-1a}
\left(D^{\mu,\nu}_{H;0+} Z\right)(t) =
- \Phi[Z(t- \varepsilon)] \cdot \sum^{\infty}_{k=1} \delta (T^{-1}t-k) ,
\end{equation}
and the initial conditions
\begin{equation} \label{InCond-1}
\lim_{t \to 0+} \left( D^{m-(1-\nu) (N-\mu)}_{RL;0+} Z \right)(t) =b_m , \quad (m=1,...,N-1) ,
\end{equation} 
\begin{equation} \label{InCond-2}
\lim_{t \to 0+} \left( I^{(1-\nu) (N-\mu)}_{RL;0+} Z \right)(t) = b_0
\end{equation} 
has the solution in the form
\begin{equation} \label{Hilfer-Zt-1}
Z(t)= \sum^{N-1}_{m=0} 
\frac{b_m \, t^{m - (N-\mu)(1-\nu)}}{\Gamma(m+1- (N-\mu)(1-\nu))} - 
\frac{T}{\Gamma (\mu)}
\sum^{n}_{k=1} (t-kT)^{\mu-1} \, \Phi [Z(kT- \varepsilon)] 
\end{equation}
for $t \in (nT,(n+1)T)$.
\end{Theorem}

\textbf{Proof.}
The integration $I^{\mu}_{RL,0+}$ of equation \eqref{FDE-1a} leads to
\begin{equation} \label{FDE-1b} 
\left(I^{\mu}_{RL,0+} D^{\mu,\nu}_{H;0+} Z\right)(t) +
\left( I^{\mu}_{RL,0+} \left(
 \Phi [Z(\tau - \varepsilon)] \sum^{\infty}_{k=1} 
\delta (T^{-1} \tau -k) \right) \right)(t) .
\end{equation} 
Applying \eqref{FTFC-Hilfer-2}, we obtain 
\[
Z(t)- \sum^{N-1}_{m=0} 
\frac{b_m \, t^{m - (N-\mu)(1-\nu)}}{\Gamma(m+1- (N-\mu)(1-\nu))} =
\]
\begin{equation} \label{FDE-2} 
- \left( I^{\mu}_{RL,0+} \left(
 \Phi [Z(\tau - \varepsilon)] \sum^{\infty}_{k=1} 
\delta (T^{-1}\tau-k) \right) \right)(t) .
\end{equation} 
Using \eqref{FTFC-Hilfer-2}, we get
\[
Z(t)- \sum^{N-1}_{m=0} 
\frac{b_m \, t^{m - (N-\mu)(1-\nu)}}{\Gamma(m+1- (N-\mu)(1-\nu))} =
\]
\begin{equation} \label{FDE-3}
- \frac{1}{\Gamma (\mu)}
\int^t_0 (t-\tau)^{\mu-1} \, \Phi [Z(\tau - \varepsilon)] 
\sum^{\infty}_{k=1} \delta (T^{-1}\tau-k) d\tau .
\end{equation} 
If $nT<t<(n+1)T$, Eq. \eqref{FDE-3} gives
\[
Z(t)= \sum^{N-1}_{m=0} 
\frac{b_m}{\Gamma(m+1-(1-\nu) (N-\mu) )} 
\, t^{m - (1-\nu) (N-\mu) } - 
\]
\begin{equation} \label{FDE-5}
\frac{1}{\Gamma (\mu)}
\sum^{n}_{k=1} \int^t_0 \Phi [Z(\tau - \varepsilon)] \,
\delta (T^{-1}\tau-k) (t-\tau)^{\mu-1} \, d\tau .
\end{equation} 
Using the property 
\begin{equation} \label{Delta-Funct}
\int^t_0 f(\tau) \, \delta (T^{-1}\tau-k) d\tau = T \, f (kT) ,
\end{equation}
which is satisfied, if $f(\tau)$ is continuous function at $\tau=kT$ and $0<kT<t$, 
equation \eqref{FDE-5} with $nT<t<(n+1)T$ takes the form
\begin{equation} \label{FDE-6}
Z(t)= \sum^{N-1}_{m=0} 
\frac{b_m \, t^{m - (N-\mu)(1-\nu)}}{\Gamma(m+1- (N-\mu)(1-\nu))} - 
\frac{T}{\Gamma (\mu)}
\sum^{n}_{k=1} \Phi [Z(kT - \varepsilon)] \, (t-kT)^{\mu-1} .
\end{equation}

$\ \ \ \Box$ \\

Further, to obtain DMM from equation \eqref{FDE-1a}, we consider expression \eqref{Hilfer-Zt-1} for the cases $\mu \in (0,1)$, $\mu>1$, and example with $\mu \in (1,2)$, separately.


\section{Map for $0<\mu<1$}

For the case $0<\mu<1$, we have the theorem.

\begin{Theorem} \label{Theorem-mu01}
For $\mu \in (0,1)$ and $Z(t) \in \Omega^{\mu}_{-1}$, equation {FDE-1a} gives the map
\[
Z_{n+1} = Z_{n} +
\frac{b_0 \, T^{- (1-\mu)(1-\nu)}}{\Gamma(1- (1-\mu)(1-\nu))} {\cal V}_{1 - (1-\mu)(1-\nu) }(n) -
\]
\begin{equation} \label{Hilfer-MAP-Z2}
\frac{T^{\mu}}{\Gamma (\mu)} \, \Phi [Z_n] 
- \frac{T^{\mu}}{\Gamma (\mu)}
\sum^{n-1}_{k=1} {\cal V}_{\mu} (n-k) \, \Phi [Z_k] ,
\end{equation}
where
\begin{equation} \label{P-n-1_LZk}
Z_{k} =\lim_{\varepsilon \rightarrow 0+} Z(Tk-\varepsilon),
\end{equation} 
and ${\cal V}_{\mu}(z)=(z+1)^{\mu-1}-z^{\mu-1}$, $z>0$, 
and $b_0$ is given by \eqref{InCond-2}.
\end{Theorem}

\textbf{Proof.}
For the case $0<\mu<1$, solution \eqref{Hilfer-Zt-1} has the form
\begin{equation} \label{Hilfer-Zt-2}
Z(t)= \frac{b_0}{\Gamma(1 - (1-\mu)(1-\nu))}
\, t^{ - (1-\mu)(1-\nu)} - 
\frac{T}{\Gamma (\mu)}
\sum^{n}_{k=1} \Phi [Z(kT- \varepsilon)] \, (t-kT)^{\mu-1} ,
\end{equation}
with $0<\mu <1$, $0<\nu<1$, and
$b_0$ is given by expressions \eqref{InCond-2}.

For $0<\mu<1$, $t=(n+1)T-\varepsilon$) and 
\begin{equation} \label{Z-n-1}
Z_{n+1}=\lim_{\varepsilon \rightarrow 0+} Z((n+1)T-\varepsilon),
\end{equation}
Eq. \eqref{Hilfer-Zt-2} is represented as 
\begin{equation} \label{Hilfer-Zt-3}
Z_{n+1} = \frac{b_0 \, T^{- (1-\mu)(1-\nu)}}{\Gamma(1 - (1-\mu)(1-\nu))} \, (n+1)^{- (1-\mu)(1-\nu)} - 
\frac{T^{\mu}}{\Gamma (\mu)}
\sum^{n}_{k=1} (n+1-k)^{\mu-1} \, \Phi [Z_k] ,
\end{equation}
where
\begin{equation} \label{Z-k-1}
Z_k=\lim_{\varepsilon \rightarrow 0+} Z(k \, T-\varepsilon) .
\end{equation}
Using expression \eqref{Hilfer-Zt-2} for the left side of the $n$th kicks ($t=nT-\varepsilon$), we get
\begin{equation} \label{Hilfer-Zt-5}
Z_{n} = \frac{b_0 \, T^{- (1-\mu)(1-\nu)}}{\Gamma(1 - (1-\mu)(1-\nu))} \, n^{- (1-\mu)(1-\nu)} - 
\frac{T^{\mu}}{\Gamma (\mu)}
\sum^{n-1}_{k=1} (n-k)^{\mu-1} \, \Phi [Z_k] .
\end{equation}
Subtracting expression \eqref{Hilfer-Zt-5} from equation \eqref{Hilfer-Zt-3}, we obtain 
\[
Z_{n+1} = Z_n +
\frac{b_0 \, T^{- (1-\mu)(1-\nu)}}{
\Gamma(1 - (1-\mu)(1-\nu))} \, 
{\cal V}_{1- (1-\mu)(1-\nu)}(n) -
\]
\begin{equation} \label{MAP-Hilf-2}
\frac{T^{\mu}}{\Gamma (\mu)} \, \Phi [Z_n] -
\frac{T^{\mu}}{\Gamma (\mu)}
\sum^{n-1}_{k=1} {\cal V}_{\mu}(n-k) \, \Phi [Z_k] ,
\end{equation}
where
\begin{equation} \label{Vam}
{\cal V}_{\mu}(z) = (z+1)^{\mu-1} -z^{\mu-1} , 
\end{equation}
and $z > 0$.
Note that the definition of the function ${\cal V}_{\mu}(z)$ is somewhat different from that used in works 
\cite{Tarasov-Zaslavsky,Tarasov-Map1,Tarasov-Map2,Springer2010},
where $V_{\mu}(z) = z^{\mu-1} -(z-1)^{\mu-1}$
(for example, see equation 18.121 in \cite{Springer2010}). 

$\ \ \ \Box$ \\

Equation \eqref{MAP-Hilf-2} can be called the universal maps with memory. 
For $\Phi [Z]=Z-r \, Z \, (1-Z)$, equation \eqref{MAP-Hilf-2} defines the logistic DMM.

Note that for $b_0=0$, map \eqref{MAP-Hilf-2} does not depend on the type parameter $\nu \in [0,1]$. \\

For more general equation
\begin{equation} \label{FDE-1a2}
\left(D^{\mu,\nu}_{H;0+} Z\right)(t) =
- \Phi[Z(t- \varepsilon),Z^{(\alpha,\beta)}(t-\varepsilon)] 
\cdot \sum^{\infty}_{k=1} \delta (T^{-1}t-k) ,
\end{equation}
where $0<\alpha<\mu$, $\beta \in [0,1]$, and
\begin{equation}
Z^{(\alpha,\beta)}(\tau) = 
(D^{\alpha,\beta}_{H;0+}Z)(\tau) ,
\end{equation}
we can give a generalization of Theorem \ref{Theorem-mu01}.

\begin{Theorem} \label{Theorem-mu01-2}
For $\mu \in (0,1)$, $\alpha \in (0,\mu)$, and $Z(t) \in \Omega^{\mu}_{-1}$, 
equation \eqref{FDE-1a2} with condition \eqref{InCond-2} gives 
\[
Z_{n+1} = Z_{n} +
\frac{b_0 \, T^{- (1-\mu)(1-\nu)}}{\Gamma(1- (1-\mu)(1-\nu))} {\cal V}_{1 - (1-\mu)(1-\nu) }(n) -
\]
\begin{equation} \label{Zab-1}
\frac{T^{\mu}}{\Gamma (\mu)} \, \Phi [Z_n,Z^{(\alpha,\beta)}_n] 
- \frac{T^{\mu}}{\Gamma (\mu)}
\sum^{n-1}_{k=1} {\cal V}_{\mu} (n-k) \, \Phi [Z_k,Z^{(\alpha,\beta)}_k] ,
\end{equation}
\[
Z^{(\alpha,\beta)}_{n+1} = Z^{(\alpha,\beta)}_{n} +
\frac{b_0 \, T^{- (1-\mu)(1-\nu)-\alpha}}{\Gamma(1- (1-\mu)(1-\nu)-\alpha)} {\cal V}_{1 - (1-\mu)(1-\nu) -\alpha}(n) -
\]
\begin{equation} \label{Zab-2}
\frac{T^{\mu}}{\Gamma (\mu)} \, \Phi [Z_k,Z^{(\alpha,\beta)}_k] 
- \frac{T^{\mu}}{\Gamma (\mu-\alpha)}
\sum^{n-1}_{k=1} {\cal V}_{\mu-\alpha} (n-k) \, \Phi [Z_k,Z^{(\alpha,\beta)}_k] ,
\end{equation}
where
\begin{equation} \label{Z-abk}
Z_{k} =\lim_{\varepsilon \rightarrow 0+} Z(Tk-\varepsilon),
Z^{(\alpha,\beta)}_{k} =\lim_{\varepsilon \rightarrow 0+} 
Z^{(\alpha,\beta)}(Tk-\varepsilon) ,
\end{equation} 
and ${\cal V}_{\mu}(z)=(z+1)^{\mu-1}-z^{\mu-1}$, $z>0$, 
and $b_0$ is given by \eqref{InCond-2}.
\end{Theorem}

\textbf{Proof.}
For the case $0<\mu<1$, problem \eqref{FDE-1a2} with \eqref{InCond-2} has the solution
\[
Z(t)= \frac{b_0}{\Gamma(1 - (1-\mu)(1-\nu))}
\, t^{ - (1-\mu)(1-\nu)} - 
\]
\begin{equation} \label{NEW-Zab}
\frac{T}{\Gamma (\mu)}
\sum^{n}_{k=1} \Phi [Z(kT- \varepsilon),Z^{(\alpha,\beta)}(Tk-\varepsilon) ] \, (t-kT)^{\mu-1} H(t-Tk),
\end{equation}
where $H(z)=1$ for $z>0$, and $H(z)=0$ for $z<0$,
which is proved similarly to Theorem \ref{Hilfer-Th-Zt-1}.
Using
\begin{equation} 
D^{\alpha,\beta}_{H;0+}
(\tau-a)^{w} H(\tau-a) = D^{\alpha,\beta}_{H;a+}
(\tau-a)^{w} =
\frac{\Gamma(w+1)}{\Gamma(w+1-\alpha)} (\tau-a)^{w-\alpha} ,
\end{equation}
for $w>-1$, we get
\[
Z^{(\alpha,\beta)}(t)= \frac{b_0}{\Gamma(1 - (1-\mu)(1-\nu)-\alpha)}
\, t^{ - (1-\mu)(1-\nu)-\alpha} - 
\]
\begin{equation} \label{NEW-Zab2}
\frac{T}{\Gamma (\mu)}
\sum^{n}_{k=1} \Phi [Z(kT- \varepsilon),Z^{\alpha,\beta}(Tk-\varepsilon) ] \, (t-kT)^{\mu-1-\alpha} ,
\end{equation}
for $tn<t<T(n+1)$.
Then similarly to Theorem \ref{Theorem-mu01}, we get
maps \eqref{Zab-1}, \eqref{Zab-2}.

$\ \ \ \Box$ \\


\section{Map for $\mu>1$}

To derive DMM from FDE \eqref{FDE-1e} of the order $\mu >1$, we should define generalized momenta and 
represent equation \eqref{FDE-1e} in the Hamiltonian form \cite[pp.409-453]{Springer2010}.

Let us define the generalized momenta
\begin{equation} \label{Pkt-1}
P_k (t) = \frac{d^k}{dt^k} 
\left( I^{(1-\nu) (N-\mu)}_{RL;0+} Z \right)(t) ,
\quad (k=1,...,N-1) ,
\end{equation} 
where $\mu>1$, $\nu \in [0,1]$, and $N-1<\mu\le N$
($N \ge 2$ for $\mu >1$). 
We can also consider the variable
\begin{equation} 
P_0(t) = I^{(1-\nu) (N-\mu)}_{RL;0+} Z(t) .
\end{equation}
Expression \eqref{Pkt-1} can used to 
define the generalized momenta in the following form.

\begin{Definition}
Let $Z(t)\in AC^{s}\left[a,b\right]$, where the space $AC^{s}\left[a,b\right]$ consists of functions $Z(t)$, which have continuous derivatives up to order $s-1$ on $\left[a,b\right]$ and function $Z^{(s-1)}(t)$ is absolutely continuous on the interval $\left[a,b\right]$.
Then the generalized momenta $P_s (t)$ are defined by the equation 
\begin{equation} \label{Pkt-k}
P_s (t) = \left(D^{s-(1-\nu) (N-\mu)}_{RL;a+}Z \right)(t) ,
\quad (s=1,...,N-1) ,
\end{equation}
where $\mu >1$ ($N \ge 2$) and
\begin{equation}
s-1<s-(1-\nu) (N-\mu)<s .
\end{equation}
\end{Definition}

For simplicity, we will assume the parameter $a$ to be zero in the definition of the generalized momenta. 

The operator $D^{\mu,\nu}_{H;0+}$ of the order $\mu \in (N-1,N)$
($N \ge 2$ for $\mu >1$) and type $\nu \in [0,1]$ can be represented in the form 
\[
\left(D^{\mu,\nu}_{H;0+} Z\right)(t) =
\left( I^{\nu (N-\mu)}_{RL;0+} \frac{d^N}{dt^N} 
I^{(1-\nu) (N-\mu)}_{RL;0+}Z \right)(t) =
\]
\[
\left( I^{\nu (N-\mu)}_{RL;0+}\frac{d^{N-k}}{dt^{N-k}}
 \right) 
\left( \frac{d^k}{dt^k}
I^{(1-\nu) (N-\mu)}_{RL;0+}Z \right)(t) =
\]
\begin{equation}
\left( D^{N-k - (N-\mu) \nu }_{C;0+} \right) 
\left( D^{k-(1-\nu) (N-\mu)}_{RL;0+}Z \right)(t) .
\end{equation}

As a result, the following theorem was proved.

\begin{Theorem}
Let $Z(t) \in \Omega^{\mu}_{-1}$, $N-1 < \mu \le N$, $N \in \mathbb{N}$ ($N \ge 2$ for $\mu >1$).
Then the equality
\begin{equation} \label{Hilfer-H-CRL}
\left(D^{\mu,\nu}_{H;0+} Z\right)(t) =
\left( D^{N-k-\nu (N-\mu)}_{C;0+} \, 
D^{k-(1-\nu) (N-\mu)}_{RL;0+}Z \right)(t) 
\end{equation} 
holds, where $k=1,...,N-1$.
The orders of the Caputo and Riemann-Liouville operators in equation \eqref{Hilfer-H-CRL} 
belong to the intervals $(N-1-k,N-k)$ and $(k-1,k)$, respectively, that is, they satisfy the inequalities 
\begin{equation}
N-k-1 < N-k-\nu (N-\mu) < N-k , \quad
k-1 < k-(1-\nu) (N-\mu) < k ,
\end{equation}
where we take into account that $0 \le \nu (N-\mu) <1$ and
$0 \le (1-\nu) (N-\mu) <1$.
\end{Theorem}

Equation \eqref{Hilfer-H-CRL} represents the HFD of the coordinate variable $Z(t)$
as the Caputo fractional derivative of the momentuma
$P_s(t)$, which are defined by expression \eqref{Pkt-k}, by the equation
\begin{equation} \label{Hilfer-H-CPt}
\left(D^{\mu,\nu}_{H;0+} Z\right)(t) =
\left( D^{N-s-\nu (N-\mu)}_{C;0+} \, P_s \right) (t) .
\end{equation}
where $N-1 < \mu \le N$, $0 \le \nu \le 1$,
$s=1,...,N-1$. 


Let us represent equation \eqref{FDE-1} with $\mu>1$ in the Hamiltonian form.

For $\mu \in (1,2)$, we have $N=2$, and equation \eqref{FDE-1} can be represented in the Hamiltonian form
\begin{equation} \label{FDE-2a}
\left(D^{1-(1-\nu) (2-\mu)}_{RL;0+}Z \right)(t) = P_1(t) ,\end{equation}
\begin{equation} \label{FDE-2b}
\left( D^{1-\nu (2-\mu)}_{C;0+} \, P_1 \right) (t)
= - \Phi[Z(t)] \sum^{\infty}_{k=1} \delta (T^{-1}t-k) .
\end{equation}

For $\mu \in (N-1,N)$, $N>2$, equation \eqref{FDE-1} gives
\begin{equation} \label{FDE-3a}
\left(D^{1-(N-\mu) (1-\nu) }_{RL;0+}Z \right)(t) = P_1(t) ,\end{equation}
\begin{equation} \label{FDE-3b}
\frac{d P_k (t)}{dt} = P_{k+1}(t) , \quad (k=1,...,N-2)
\end{equation}
\begin{equation} \label{FDE-3c}
\left( D^{1-\nu (N-\mu)}_{C;0+} \, P_{N-1} \right) (t)
= - \Phi[Z(t)] \sum^{\infty}_{k=1} \delta (T^{-1}t-k) .
\end{equation}


To solve these equations with $\mu>1$ for generalized momenta,
we can use an analog of Theorem \ref{Hilfer-Th-Zt-1}, which is formulated thought the generalized momenta $P_s(t)$, where $s=1,...,N-1$.

Using equality \eqref{Hilfer-H-CPt}, Eq. \eqref{FDE-1e} can be rewritten in the form
\begin{equation} \label{FDE-1-Ps}
\left( D^{N-s-\nu (N-\mu)}_{C;0+} \, P_s \right) (t) =
- \Phi[Z(t- \varepsilon)] \sum^{\infty}_{k=1} \delta (T^{-1}t-k).
\end{equation}
Using \eqref{Pkt-k}, coefficients \eqref{p-k0-1}, which characterize the initial condition, can be represented through the generalized momenta in the form
\begin{equation} \label{p-k0-2}
p^{\mu,\nu}_s(0+) = \lim_{t \to 0+} P_s(t) , \quad
(s=0,...,N-1) .
\end{equation} 

As a result, we can formulate the theorem for equation \eqref{FDE-1-Ps}.

\begin{Theorem} \label{Hilfer-Th-Ps}
Let $P_s(t)\in AC^{N-s}[0,b]$ or 
$P_s(t)\in C^{N-s}[0,b]$, $N-1 <\mu \le N$. 
The Cauchy problem for the fractional differential equation
\eqref{FDE-1-Ps} of the order $N-s-\nu (N-\mu)$
and the initial conditions
\begin{equation} 
\lim_{t \to 0+} P_s(t) = b_s , \quad (s=1,...,N-1) ,
\end{equation} 
and condition \eqref{InCond-2}
has the solution for $Tn < t<T(n+1)$ in the form
\[
P_s(t)= \sum^{N-s-1}_{m=0} \frac{b_m}{m!} P^{(m)}_s(0+) - 
\]
\begin{equation} \label{Hilfer-Map-Ps}
\frac{T}{\Gamma ((N-s)-\nu (N-\mu))}
\sum^{n}_{k=1} (t-kT)^{(N-s)-\nu (N-\mu)-1} \, 
\Phi [Z(kT- \varepsilon)] ,
\end{equation}
where $s=1,...,N-1$, $N-1<\mu \le N$, $0 \le \nu \le 1$ and
\[
N-s-1 < (N-s)-\nu (N-\mu) <N-s .
\]
\end{Theorem}

{\bf Proof.}
Equation \eqref{FDE-1-Ps} can be rewritten in the form
\begin{equation} \label{FDE-1-PsNew}
\left( D^{\mu^{\prime}}_{C;0+} \, P_s \right) (t) =
- \Phi[Z(t- \varepsilon)] \sum^{\infty}_{k=1} \delta (T^{-1}t-k) ,
\end{equation}
where
\begin{equation}
N^{\prime} =N-s \quad ,
\mu^{\prime} = N-s-\nu (N-\mu) .
\end{equation}
The action the Riemann-Liouvill integral on equation
\eqref{FDE-1-PsNew} gives
\begin{equation} 
\left( I^{\mu^{\prime}}_{RL;0+} D^{\mu^{\prime}}_{C;0+} \, P_s \right) (t) =
- \left( I^{\mu^{\prime}}_{RL;0+} 
\Phi [Z(t- \varepsilon)] \sum^{\infty}_{k=1} \delta (T^{-1}t-k) \right) (t) .
\end{equation}
Then we use the second fundamental theorem in the form of Lemma 2.22 \cite[p.~96]{FC4}. 
In our case this theorem can be formulated in the following form: 
Let $N^{\prime}-1 <\mu^{\prime} < N^{\prime}$. 
If $P_s(t)\in AC^{N^{\prime}}[0,b]$ or 
$P_s(t)\in C^{N^{\prime}}[0,b]$, then 
\begin{equation} \label{EQ5-4_28} 
\left(I^{\mu^{\prime}}_{RL;0+} D^{\mu^{\prime}}_{C;0+} 
P_s \right)(t) =
P_s(t)- \sum^{N^{\prime}-1}_{m=0} p^C_m (0+) \, t^m , 
\end{equation} 
where
\begin{equation} \label{EQ5-4_29} 
p^C_m (a+)= \lim_{t \to a+} \frac{P_s^{(m)}(t)}{m!}. 
\end{equation} 
In this proof, we should use $N^{\prime}=N-s$,
and $\mu^{\prime}=(N-s)-\nu (N-\mu)$.

Then, to get maps \eqref{Hilfer-Map-Ps}, we realize
transformation analogous to transformations in the proof of Theorem \eqref{Hilfer-Th-Zt-1}.

$\ \ \ \Box$ \\


\section{Example: Map for $1<\mu <2$}

For $\mu \in (1,2)$, we have $N=2$ and equation \eqref{FDE-1} can be represented in the Hamiltonian form
\begin{equation} \label{FDE-2aN}
\left(D^{1-(1-\nu) (2-\mu)}_{RL;0+}Z \right)(t) = P_1(t) ,
\end{equation}
\begin{equation} \label{FDE-2bN}
\left( D^{1-\nu (2-\mu)}_{C;0+} \, P_1 \right) (t)
= - \Phi[Z(t-\varepsilon)] \sum^{\infty}_{k=1} 
\delta (T^{-1}t-k) .
\end{equation}
For these equations, we can formulate the following two Lemmas:

\begin{Lemma} \label{Lemma-P}
Let $0<1- (2-\mu) \nu <1$. If $P_1(t)\in AC^{1}[0,b]$ or 
$Z(t)\in C^{1}[0,b]$, 
then equation \eqref{FDE-2bN} leads to the map
\begin{equation} \label{FDE-2bN9b}
P_{1,n+1} = P_{1,n} 
- \frac{ T^{1 - \nu (2-\mu)}}{\Gamma (1 - \nu (2-\mu))} \, \Phi [Z_n] -
\frac{ T^{ 1-\nu(2-\mu) }}{\Gamma ( 1-\nu(2-\mu) )} 
\sum^{n-1}_{k=1} {\cal V}_{1 - \nu (2-\mu)}(n-k) \, \Phi [Z_k] ,
\end{equation}
where 
\begin{equation} \label{P-n-1-LP}
P_{1,n} =\lim_{\varepsilon \rightarrow 0+} P_1(Tn-\varepsilon),
\end{equation}
and ${\cal V}_{\mu}(z)$ is given in \eqref{Vam}.
\end{Lemma}

\textbf{Proof.}
Let us write equation \eqref{FDE-2bN} as
\begin{equation} \label{FDE-2bN2}
\left( D^{\mu^{\prime}}_{C;0+} \, P_1 \right) (t) = - 
\Phi[Z(t-\varepsilon)] \sum^{\infty}_{k=1} \delta (T^{-1}t-k) ,
\end{equation}
where
\[
\mu^{\prime} = 1-\nu (2-\mu) , \quad 
0<\mu^{\prime} <1 .
\]
Using Lemma 2.22 of \cite[p.~96]{FC4}) for $\mu^{\prime} \in (0,1)$, $N^{\prime}=1$, in the form
\begin{equation} \label{FTFC-RL-3}
\left(I^{\mu^{\prime} }_{RL;0+} D^{\mu^{\prime}}_{C;0+} P_1 \right)(t) = P_1(t)- P_1(0+) , 
\end{equation} 
where
\begin{equation} \label{FTFC-RL-3p}
P_1(0+) = \lim_{t \to 0+} \, P_1(t) . 
\end{equation} 
Using equality \eqref{FTFC-RL-3} and equation \eqref{FTFC-Hilfer-2} in the form
\begin{equation} \label{EQ1-27prime} 
\left(I^{\mu^{\prime}}_{RL;0+} Z \right)(t) = 
\int^t_0 \frac{1}{\Gamma (\mu^{\prime}) (t-\tau)^{1- \mu^{\prime}}} Z (\tau) d\tau , 
\end{equation} 
equation \eqref{FDE-2bN2} gives
\begin{equation} \label{FDE-2bN3}
P_1(t) - P_1(0+) =
- \frac{1}{\Gamma (\mu^{\prime} )}
\int^t_0 
\left( \Phi [Z(\tau-\varepsilon)] \sum^{\infty}_{k=2} \delta (T^{-1}\tau-k) \right) \, (t-\tau)^{\mu^{\prime}-1} \, d\tau .
\end{equation}

For $nT<t<(n+1)T$, equation \eqref{FDE-2bN3} takes the form
\begin{equation} \label{FDE-2bN5}
P_1(t) - P_1(0+) = - \sum^{n}_{k=1} \int^t_0 
\frac{1}{\Gamma (\mu^{\prime}) (t-\tau)^{1-\mu^{\prime}} } 
\left( \Phi [Z(\tau-\varepsilon)] \delta (T^{-1}\tau-k)\right) \, d\tau .
\end{equation} 
Using \eqref{Delta-Funct} for $\tau=kT$ and $0<kT<t$, equation \eqref{FDE-2bN5} gives
\begin{equation} \label{FDE-2bN6}
P_1(t) = P_1(0+) - 
\frac{ T}{\Gamma (\mu^{\prime} )} 
\sum^{n}_{k=1}\Phi [Z(kT-\varepsilon)] 
(t-kT)^{\mu^{\prime}-1} .
\end{equation} 
For $t=T(n+1)-\varepsilon$, and
\begin{equation} \label{P-n-1}
P_{1,n+1} =\lim_{\varepsilon \rightarrow 0+} P_1((n+1)T-\varepsilon) ,
\end{equation}
equation \eqref{FDE-2bN6} is 
\begin{equation} \label{FDE-2bN7}
P_{1,n+1} = P_{1,0} - 
\frac{ T^{\mu^{\prime}}}{\Gamma (\mu^{\prime})} 
\sum^{n}_{k=1} \Phi [Z_k] \, (n+1-k)^{\mu^{\prime}-1} .
\end{equation} 
Expression \eqref{FDE-2bN7} for $t=Tn-\varepsilon$ is
\begin{equation} \label{FDE-2bN8}
P_{1,n} = P_{1,0} - 
\frac{ T^{\mu^{\prime}}}{\Gamma (\mu^{\prime})} 
\sum^{n-1}_{k=1} (n-k)^{\mu^{\prime}-1} \, \Phi [Z_k] .
\end{equation}
Subtracting expression \eqref{FDE-2bN8} from equation \eqref{FDE-2bN7}, we obtain 
\begin{equation} \label{FDE-2bN9}
P_{1,n+1} - P_{1,n} = 
- \frac{ T^{\mu^{\prime}}}{\Gamma (\mu^{\prime})} 
\, \Phi [Z_n]
- \frac{ T^{\mu^{\prime}}}{\Gamma (\mu^{\prime})} 
\sum^{n-1}_{k=1} 
\left( (n+1-k)^{\mu^{\prime}-1} - (n-k)^{\mu^{\prime}-1} 
\right) \, \Phi [Z_k] ,
\end{equation} 
where $\mu^{\prime} = 1 - (2-\mu) \nu $.

Using function \eqref{Vam}, equation \eqref{FDE-2bN9} takes form \eqref{FDE-2bN9b}. 

$\ \ \ \Box$ \\


\begin{Lemma} \label{Lemma-Z}
Let $0< 1- (1-\nu) (2-\mu) <1$. If $Z(t)\in L_{1}(a,b)$ 
has a summable fractional derivative 
$\left(D^{1-(1-\nu) (2-\mu)}_{RL;a+}Z\right)(t)$, i.e., 
$\left(I^{ (1-\nu) (2-\mu) }_{RL;a+}Z\right)(t)\in AC^{1}[0,b]$, then equation \eqref{FDE-2aN} gives the map
\[
Z_{n+1} = Z_{n} +
\frac{b_0 \, T^{- (2-\mu)(1-\nu)}}{\Gamma(1- (2-\mu)(1-\nu))} {\cal V}_{1 - (2-\mu)(1-\nu) }(n) +
\]
\[
\frac{b_1 \, T^{1 - (1-\nu) (2-\mu) }}{\Gamma(2- (1-\nu) (2-\mu) )} {\cal V}_{2 - (1-\nu) (2-\mu)}(n) 
\]
\begin{equation} 
- \frac{T^{\mu}}{\Gamma (\mu)} \, \Phi [Z_n] 
- \frac{T^{\mu}}{\Gamma (\mu)}
\sum^{n-1}_{k=1} {\cal V}_{\mu} (n-k) \, \Phi [Z_k] ,
\end{equation}
where
\begin{equation} \label{P-n-1_LZ}
Z_{n} =\lim_{\varepsilon \rightarrow 0+} Z(Tn-\varepsilon),
\end{equation} 
and ${\cal V}_{\mu}(z)$ is defined by equation \eqref{Vam},
and $b_0$, $b_1$ are defined by equations \eqref{InCond-1}, \eqref{InCond-2}.
\end{Lemma}

\textbf{Proof.}
Using expression \eqref{Hilfer-Zt-1}, the solution of FDE \eqref{FDE-2aN} for $Tn <t<T(n+1)$ is
\[
Z(t)= \frac{b_0 \, t^{- (2-\mu)(1-\nu)}}{\Gamma(1- (1-\nu)(2-\mu) )} +
\frac{b_1 \, t^{1 - (1-\nu) (2-\mu) }}{\Gamma(2- (1-\nu) (2-\mu) )} - 
\]
\begin{equation} \label{Hilfer-Zt-1-2}
\frac{T}{\Gamma (\mu)}
\sum^{n}_{k=1} \Phi [Z(kT-\varepsilon)] (t-kT)^{\mu-1} ,
\end{equation}
where $1<\mu < 2$, $0<\nu<1$, and
$b_1$ is given by expression \eqref{InCond-1} and 
$b_0$ is defined by \eqref{InCond-2}.

Expression \eqref{Hilfer-Zt-1-2} with $t=T(n+1)-\varepsilon$ gives
\[
Z_{n+1} = \frac{b_0 \, (T(n+1))^{- (1-\nu) (2-\mu) }}{\Gamma(1- (1-\nu) (2-\mu) )} +
 \frac{b_1 \, (T(n+1))^{1 - (1-\nu) (2-\mu) } }{\Gamma(2- (1-\nu)(2-\mu) )} - 
\]
\begin{equation} \label{Hilfer-Zt-1-2n1}
\frac{T^{\mu}}{\Gamma (\mu)}
\sum^{n}_{k=1} \Phi [Z_k] \, (n+1-k)^{\mu-1} .
\end{equation}
\[
Z_{n} = \frac{b_0 \, (nT)^{- (2-\mu)(1-\nu)}}{\Gamma(1- (1-\nu) (2-\mu) )} +
\frac{b_1 \, (nT)^{1 - (2-\mu)(1-\nu)}}{\Gamma(2- (2-\mu)(1-\nu))} - 
\]
\begin{equation} \label{Hilfer-Zt-1-2n2}
\frac{T^{\mu}}{\Gamma (\mu)}
\sum^{n-1}_{k=1} (n-k)^{\mu-1} \, \Phi [Z_k] .
\end{equation}
Subtracting expression \eqref{Hilfer-Zt-1-2n2} from equation 
\eqref{Hilfer-Zt-1-2n1}, we get
\[
Z_{n+1} - Z_{n}= \frac{b_0 \, T^{- (1-\nu) (2-\mu) }}{\Gamma(1- (2-\mu)(1-\nu))} {\cal V}_{1 - (1-\nu) (2-\mu) }(n) +
\]
\[
\frac{b_1 \, T^{1 - (2-\mu)(1-\nu)}}{\Gamma(2- (2-\mu)(1-\nu))} {\cal V}_{2 - (2-\mu)(1-\nu) }(n) -
\]
\begin{equation} \label{Hilfer-MAP-Z}
\frac{T^{\mu}}{\Gamma (\mu)} \, \Phi [Z_n] -
\frac{T^{\mu}}{\Gamma (\mu)}
\sum^{n-1}_{k=1} {\cal V}_{\mu}(n-k) \, \Phi [Z_k] .
\end{equation}

$\ \ \ \Box$ \\


As a result, using Lemma \ref{Lemma-Z} and Lemma \ref{Lemma-P}, we get the following theorem.

\begin{Theorem} \label{Hilfer-Th-Zt-2}
Equations \eqref{FDE-2aN}, \eqref{FDE-2bN} lead to the map
\[
Z_{n+1} = Z_{n} +
\frac{b_0 \, T^{- (2-\mu)(1-\nu)}}{\Gamma(1- (1-\nu) (2-\mu) )} {\cal V}_{1 - (1-\nu) (2-\mu)}(n) +
\]
\[
 \frac{b_1 \, T^{1 - (2-\mu)(1-\nu)}}{\Gamma(2- (2-\mu)(1-\nu))} {\cal V}_{2 - (2-\mu)(1-\nu) }(n) -
\]
\begin{equation} \label{Hilfer-MAP-Z2b}
\frac{T^{\mu}}{\Gamma (\mu)} \, \Phi [Z_n] -
\frac{T^{\mu}}{\Gamma (\mu)}
\sum^{n-1}_{k=1} {\cal V}_{\mu}(n-k) \, \Phi [Z_k] ,
\end{equation}
and
\begin{equation} \label{Hilfer-MAP-P}
P_{1,n+1} = P_{1,n} 
- \frac{ T^{1 - \nu (2-\mu)}}{\Gamma (1 - \nu (2-\mu))} \, \Phi [Z_n] -
\frac{ T^{ 1-\nu(2-\mu) }}{\Gamma ( 1-\nu(2-\mu) )} 
\sum^{n-1}_{k=1} {\cal V}_{1 - \nu (2-\mu)}(n-k) \, \Phi [Z_k] .
\end{equation} 
\end{Theorem}

\textbf{Proof.}
Using equations \eqref{FDE-2bN9b}, \eqref{P-n-1_LZ} that were proved in Lemma \ref{Lemma-Z} and Lemma \ref{Lemma-P}, we get equations \eqref{Hilfer-MAP-P} and \eqref{Hilfer-MAP-Z2b} that describe the DMM for $\mu \in (1,2)$.

$\ \ \ \Box$ \\

Similarly, one can prove a generalization of Theorem \ref{Hilfer-Th-Zt-2}.

\begin{Theorem} 
Equations 
\begin{equation} 
\left(D^{1-(1-\nu) (2-\mu)}_{RL;0+}Z \right)(t) = P_1(t) ,
\end{equation}
\begin{equation} \label{FDE-2bN-new}
\left( D^{1-\nu (2-\mu)}_{C;0+} \, P_1 \right) (t)
= - \Phi[Z(t-\varepsilon),P_1(t-\varepsilon)] 
\sum^{\infty}_{k=1} \delta (T^{-1}t-k) 
\end{equation}
with $\mu \in (1,2)$ lead to the map
\[
Z_{n+1} = Z_{n} +
\frac{b_0 \, T^{- (2-\mu)(1-\nu)}}{\Gamma(1- (1-\nu) (2-\mu) )} {\cal V}_{1 - (1-\nu) (2-\mu)}(n) +
\]
\[
 \frac{b_1 \, T^{1 - (2-\mu)(1-\nu)}}{\Gamma(2- (2-\mu)(1-\nu))} {\cal V}_{2 - (2-\mu)(1-\nu) }(n) -
\]
\begin{equation} 
\frac{T^{\mu}}{\Gamma (\mu)} \, \Phi [Z_n,P_{1,n}] -
\frac{T^{\mu}}{\Gamma (\mu)}
\sum^{n-1}_{k=1} {\cal V}_{\mu}(n-k) \, \Phi [Z_k,P_{1,k}] ,
\end{equation}
and
\[
P_{1,n+1} = P_{1,n} 
- \frac{ T^{1 - \nu (2-\mu)}}{\Gamma (1 - \nu (2-\mu))} \, \Phi [Z_n,P_{1,n}] -
\]
\begin{equation} \label{Hilfer-MAP-P-new}
\frac{ T^{ 1-\nu(2-\mu) }}{\Gamma ( 1-\nu(2-\mu) )} 
\sum^{n-1}_{k=1} {\cal V}_{1 - \nu (2-\mu)}(n-k) \, \Phi [Z_k,P_{1,k}] .
\end{equation} 
\end{Theorem}

It should be emphasized that the proposed DMM are derived from the FDE without any approximations.

Equations \eqref{Hilfer-MAP-Z2b} and \eqref{Hilfer-MAP-P} can be called the universal maps with memory and $\mu >1$.

For $\Phi [Z]=-Z$, equations \eqref{Hilfer-MAP-Z2b} and \eqref{Hilfer-MAP-P} give a generalization of the Anosov type dynamical system.

For $\Phi [Z]=\sin (Z)$, equations \eqref{Hilfer-MAP-Z2b} and \eqref{Hilfer-MAP-P} are a generalization of the standard or
Chirikov-Taylor map \cite{Chirikov}.


\section{Map for momentum $P_s (t)=Z^{(s)}(t)$}

In general, we can also consider the standard momenta 
\begin{equation} \label{P-New-k}
P_s (t) = \frac{d^s Z(t)}{dt^s} , \quad (s=1,...,N-1) .
\end{equation} 
Then we can use 
\begin{equation} \label{P-New-k2}
\left( t^{w} \right)^{(s)} := \frac{d^s}{dt^s} t^{w} = 
\frac{\Gamma(w+1)}{\Gamma (w -s+1)} \, t^{w -s} ,
\end{equation}
where $s \in \mathbb{N}$, $w=k - (N-\mu)(1-\nu)$ to get
the derivatives of the integer order $s>0$:
\begin{equation} \label{P-New-k2a}
\left( t^{k - (N-\mu)(1-\nu)}\right)^{(s)} =
\frac{\Gamma(k - (N-\mu)(1-\nu)+1)}{\Gamma (k - (N-\mu)(1-\nu) -s+1)} t^{k - (N-\mu)(1-\nu)-s}
\end{equation}
and
\begin{equation} \label{P-New-k2b}
\left( (t-kT)^{\mu-1} \right)^{(s)} =
\frac{\Gamma(\mu)}{\Gamma (\mu -s)}
(t-kT)^{\mu-1-s} 
\end{equation}
for $t>kT>0$.

Using expression \eqref{Hilfer-Zt-1} and equations \eqref{P-New-k2a}, we get
\[
P_s (t) = \sum^{N-1}_{m=0} 
\frac{b_m}{\Gamma (m - (1-\nu) (N-\mu) -s+1)} 
t^{m - (1-\nu) (N-\mu) - s} - 
\]
\begin{equation} \label{Hilfer-Zt-k}
\frac{T}{\Gamma (\mu -s)}
\sum^{n}_{k=1} \Phi [Z(kT - \varepsilon)] (t-kT)^{\mu-1-s} ,
\end{equation}
where $s=1,...,N-1$.

For $\mu \in (N-1,N)$, $N \in \mathbb{N}$, $0<\nu<1$, 
the DMM is described by the equations
\[
P_{s,n+1}-P_{s,n} = \sum^{N-1}_{k=0} 
\frac{b_m \, T^{m - (1-\nu) (N-\mu) -s}}{\Gamma (m - (1-\nu)(N-\mu) -s+1)} 
{\cal V}_{m - (1-\nu)(N-\mu) -s+1}(n) - 
\]
\begin{equation} \label{Hilfer-Zt-k-2}
\frac{T^{\mu-s}}{\Gamma (\mu -s)} \Phi [Z_n]
- \frac{T^{\mu-s}}{\Gamma (\mu -s)}
\sum^{n-1}_{k=1} {\cal V}_{\mu-1-s}(n-k) \, \Phi [Z_k] ,
\end{equation}
where $s=0,...,N-1$, $P_{0,n}=Z_n$, and $b_m$ is given by expressions \eqref{InCond-1} and \eqref{InCond-2}.


For $\mu \in (1,2)$, the standard momentum is defined by the equation 
\begin{equation} \label{P-New-1c}
P_1 (t) = \frac{d Z(t)}{dt} .
\end{equation} 

Equation \eqref{Hilfer-Zt-k} gives
\[
P_1 (t)= \frac{b_0 \, 
t^{- (2-\mu)(1-\nu)-1}}{\Gamma(- (2-\mu)(1-\nu))} +
\frac{b_1 \, t^{ - (2-\mu)(1-\nu)}}{\Gamma(1- (2-\mu)(1-\nu))} - 
\]
\begin{equation} \label{Hilfer-Zt-1-2d}
\frac{T}{\Gamma (\mu-1)}
\sum^{n}_{k=1} (t-kT)^{\mu-2} \, \Phi [Z(kT)] ,
\end{equation}
where we use $\Gamma(1+z) = z \, \Gamma (z)$.

In the case $\mu \in (1,2)$, the DMM is 
\[
Z_{n+1} = Z_{n} +
\frac{b_0 \, T^{- (1-\nu)(2-\mu) }}{\Gamma(1- (2-\mu)(1-\nu))} {\cal V}_{1 - (2-\mu)(1-\nu) }(n) +
\]
\[
 \frac{b_1 \, T^{1 - (2-\mu)(1-\nu)}}{\Gamma(2- (1-\nu) (2-\mu) )} {\cal V}_{\mu}(n) -
\]
\begin{equation} \label{Hilfer-Map-B-1}
\frac{T^{\mu}}{\Gamma (\mu)} \, \Phi [Z_n] -
\frac{T^{\mu}}{\Gamma (\mu)}
\sum^{n-1}_{k=1} {\cal V}_{ 1- (1-\nu) (2-\mu)}(n-k) \, \Phi [Z_k] .
\end{equation}
\[
P_{1,n+1} -P_{1,n} = \frac{b_0 \, 
T^{- (2-\mu)(1-\nu)-1}}{\Gamma(- (1-\nu) (2-\mu) )} 
{\cal V}_{- (1-\nu) (2-\mu) }(n) +
\]
\[
\frac{b_1 \, T^{ - (1-\nu) (2-\mu) }}{\Gamma(1- (2-\mu)(1-\nu))} {\cal V}_{1- (1-\nu)(2-\mu) } (n) - 
\]
\begin{equation} \label{Hilfer-Map-B-2}
\frac{T^{\mu-1}}{\Gamma (\mu-1)} \, \Phi [Z_n] -
\frac{T^{\mu-1}}{\Gamma (\mu-1)}
\sum^{n-1}_{k=1} {\cal V}_{\mu-1}(n-k) \, \Phi [Z_k] .
\end{equation}

It should be emphasized that the proposed DMM are derived from the FDE with periodic kicks without any approximations.

Equations \eqref{Hilfer-Map-B-1} and \eqref{Hilfer-Map-B-2} describe the universal DMM.
For $\Phi [Z]=-Z$, equations \eqref{Hilfer-Map-B-1} and \eqref{Hilfer-Map-B-2} describe the Anosov DMM.
For $\Phi [Z]=\sin (Z)$, equations \eqref{Hilfer-Map-B-1} and \eqref{Hilfer-Map-B-2} give the Chirikov-Taylor DMM.

For $\nu=0$ and $\nu=1$, we can get the DMM that have been proposed in works \cite{Tarasov-Zaslavsky,Tarasov-Map1,Tarasov-Map2} 
(see also \cite[pp.409-453]{Springer2010} and 
\cite{Tarasov-Map3,TT-Logistic}), where they were obtained from FDE with the Caputo and Riemann-Liouville operators.


\section{Conclusion}

In this paper, new DMM are derived from equations with Hilfer derivative of non-integer and periodic kicks.
We obtain the exact solution of the proposed nonlinear FDE with kicks.
The proposed discrete maps with memory are derived from these solutions without any approximations.
 
The proposed DMM can be used to interpolate smoothly between the discrete maps that are derived from FDE with the Riemann-Liouville ($\nu=0$) and Caputo ($\nu=1$) operators.
It is interesting to study how strange attractors and chaotic behavior for such maps change, when the parameter $\nu$ changes from zero to one. 
However, this modeling remains an open question that can be solved in future research.



\end{document}